\documentclass[aps,prd,preprint]{revtex4}
\usepackage{amssymb,amsmath,graphicx}
\begin{document}

\title{Decoherence and Recoherence in Model Quantum Systems}

\author{Jen-Tsung Hsiang}

\address{Department of Physics, National Dong Hwa University \\
Hualien, Taiwan, R.O.C.}

\author{L. H. Ford}

\address{Institute of Cosmology, Department of Physics and Astronomy \\ 
Tufts University, Medford, MA 02155 USA}

\begin{abstract}
We discuss the various manifestations of quantum decoherence in the
forms of dephasing, entanglement with the environment, and revelation
of ``which-path'' information. As a specific example, we consider
an electron interference experiment. The coupling of the coherent
electrons to the quantized electromagnetic field illustrates all of
these versions of decoherence. This decoherence has two equivalent
interpretations, in terms of photon emission or in terms of
Aharonov-Bohm phase fluctuations. We consider the case when the
coherent electrons are coupled to photons in a squeezed vacuum state.
The time-averaged result is increased decoherence. However, if only
electrons which are emitted during selected periods are counted, the
decoherence can be suppressed below the level for the photon vacuum.
This is the phenomenon of {\it recoherence}. This effect is closely related
to the quantum violations of the weak energy condition, and is restricted
by similar inequalities. We give some estimates of the magnitude of the
recoherence effect and discuss prospects for observing it
 in an electron interferometry experiment.
\end{abstract}
\maketitle
\baselineskip=14pt
\section{Introduction}

Decoherence is a nearly universal effect in quantum systems, and one
which plays a central role in the quantum to classical transition.
However, the word ``decoherence'' is often used to denote several
related, but distinct concepts. The first is dephasing, the loss of
definite phase relations between different components in a superposition
state. These phase relationships are essential for quantum
interference, which can be regarded as the key phenomenon which
distinguishes the quantum world from the classical world. In general,
interaction of a quantum system with its environment can lead to
dephasing. The second, closely related concept, is that of entanglement
between the quantum system and its environment. Entanglement by itself
need not lead to decoherence, but often the variables of the
environment are too numerous or complex to be readily measured. It is
when one gives up on any attempt to keep track of the environmental
degrees of freedom that decohenece arises. The third concept
associated with decoherence is the revelation of ``which path''
information. Any measurement in a double slit interference experiment
which reveals the path taken by the particles will destroy the
interference pattern.

\begin{figure}
\begin{center}
    \scalebox{0.5}{
        \includegraphics{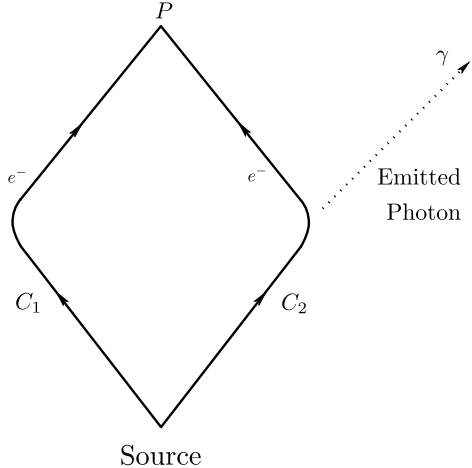}}
\caption{An electron interference experiment in which the electrons
  may take either one of two paths, $C_1$ or $C_2$, from the source to
  the point $P$ where the interference pattern
is formed. The emission of photons by the electrons tends to cause
decoherence. The detection
of an emitted photon with wavelength smaller than the path separation
can reveal which path
a particular electron takes, and hence causes
decoherence. }
\label{fig:paths}
\end{center}
\end{figure}

Here we will consider a specific example of a quantum system, coherent
electrons coupled to the quantized electromagnetic field. An electron
interference experiment, such as that illustrated in
Fig.~\ref{fig:paths}, deals with  one of the simplest possible quantum 
systems, but also one in which many beautiful experiments have been performed 
in recent years\cite{aT1982,TEMKE1989,NH93,T05,SH07}. 
The coupling to the quantized electromagnetic field allows
for photon emission by the electrons, leading to decoherence. This example
illustrates all three of the versions of decoherence discussed above.
As will be shown in Sect.~\ref{sec:AB}, the coupling to the quantized
 electromagnetic 
field produces dephasing, or a loss of contrast in the interference
pattern as a result of Aharonov-Bohm phase fluctuations. The emission
of photons leads to entanglement between the quantum state of the electrons
and that of the photons. Finally, photon emission is capable of revealing
``which path'' information. If the wavelength of an emitted photon is
less than the separation of the two electron paths in Fig.~\ref{fig:paths},
then detection of that photon reveals the path taken by a given electron.
In the cleanest version of the electron interference experiment, the
flux of electrons may be made so low that only one electron is in the 
interferometer at any one time.   

\section{Aharonov-Bohm Phase Fluctuations}
\label{sec:AB}

 Consider a double slit interference experiment in which coherent electrons
can take either one of two paths, as illustrated in Fig.~\ref{fig:paths}. 
First consider the  case of no field fluctuations.  If the amplitudes
for the electrons to take path $C_1$ and $C_2$ are $\psi_1$ and $\psi_2$,
 respectively,
 to point $P$, then the mean number of electrons at $P$ will be proportional to
\begin{equation}
n(P)= |\psi_1+\psi_2|^2 = |\psi_1|^2+|\psi_2|^2 + 
2 {\rm Re}(\psi_1\, \psi_2^*) \,.
\end{equation}  
In the presence of a classical, non-fluctuating electromagnetic field 
described by vector 
potential $A^\mu$, there will be an Aharonov-Bohm phase shift of the 
form\cite{AB1959}. 
\begin{equation}
\varphi_{AB} = e \oint_C  d x^\mu \, A_\mu \,,
\end{equation}
where the integral is taken around the closed path $C= C_1 -C_2$. Here
$e$ is the electron's charge. We will use  Lorentz-Heaviside units with
 $\hbar = c =1$. 
The phase shift alters the locations
of the interference minima and maxima, but does not alter their relative 
amplitudes, the contrast. 

If the electromagnetic field undergoes fluctuations, then the situation is 
different. In the case of Gaussian fluctuations, the fluctuating  
Aharonov-Bohm phase causes a change in the contrast by a factor of
\begin{equation}
\Gamma =  {\rm e}^W \,,    \label{E:gamma}
\end{equation}
where we define the {\it coherence functional} by
\begin{equation}
W =-\frac{1}{2} \langle \varphi_{AB}^2 \rangle 
\end{equation}
with the angular brackets denoting averaging over the fluctuations. 
This functional can be expressed as
\begin{equation}
    W=-2\pi\alpha\oint_Cdx_{\mu}\oint_Cdx'_{\nu}\;D^{\mu\nu}(x,x')\,,
\label{eq:W}
\end{equation}
where $\alpha$ is the fine-structure constant and
\begin{equation}\label{E:hada}
    D^{\mu\nu}(x,x')=\frac{1}{2}\,\bigl\langle \bigl\{A^{\mu}(x),A^{\nu}(x')
          \bigr\} \bigr\rangle\,.
\end{equation}
If the initial quantum state of the electromagnetic field is the vacuum, then
$ D^{\mu\nu}(x,x')$ is the vacuum Hadamard function. Note that the Hadamard
function itself is gauge dependent, but that $W$ is gauge invariant. 

In general, $W<0$, leading to a reduction in contrast or dephasing. At the 
same time, the coherence functional reflects the effects of photon emission.
If the electromagnetic field is initially in its vacuum state, $|0\rangle$,
 then after an electron traverses path $C_j$, it will be in the state
\begin{equation}
{\rm e}^{i e \int_{C_j}  d x^\mu \, A_\mu} \,|0\rangle \, ,
\end{equation}
which is a superposition of photon number eigenstates. Thus the descriptions
in terms of a fluctuating Aharonov-Bohm phase or in terms of photon emission
are complementary viewpoints. 

Various aspects of decoherence associated with
fluctuating or time-dependent electromagnetic fields have been discussed by
numerous authors in recent years
~\cite{SAI1990,SAI1991,lF1993,lF1994,lF1997,APZ97,VS98,BP2001,MPV2003,jH2004,jH2004a,LMV05,HL06,Mach06,HL08,AM08,HF08}. 
The connection between Aharonov-Bohm phase fluctuations and decoherence
was  discussed in by Stern  {\it et al}\cite{SAI1990}.
 The effects of boundaries,
such as a plane mirror, were considered by Ford\cite{lF1993,lF1997}.
Some of these estimates for the magnitude of the effects 
were criticized by Mazzitelli {\it et al}~\cite{MPV2003} as being 
too large.  The discrepancy is likely due to the sharp corners in the
electron paths used\cite{lF1993,lF1997}; when smooth paths are
employed\cite{MPV2003}, the effects are much smaller. This
can be understood because sharp boundaries cause large amounts of photon 
emission, which can in turn be modified by a perfectly reflecting boundary.
A different effect was discussed by Anglin {\it et al}\cite{APZ97} and
Machnikowski\cite{Mach06}, who considered the effect
of an imperfect conductor. Here the motion of electrons over the conductor 
can cause excitations inside the metal, resulting in decoherence. This
type of decoherence was recently observed by  Sonnentag and
Hasselbach\cite{SH07}. 

The initial quantum state of the electromagnetic field need not be the 
vacuum. The case of a thermal state, and the resulting increased decoherence,
was discussed by Breuer and Petruccione\cite{BP2001}. Another possibility is 
a time-dependent classical electromagnetic field\cite{jH2004,jH2004a}.
Here one really has Aharonov-Bohm phase variations rather than fluctuations.
However in most experiments, one would average over the emission time of
the electrons and effectively lose the information carried by this 
time-dependent phase, leading to decoherence.

\section{Squeezed States and Recoherence}

Now we consider the situation when the initial state of the photon field is
not the vacuum, but rather a squeezed state\cite{Stoler,Caves}. 
Such states have the property 
that they can temporarily suppress the quantum fluctuations in a given
variable below the vacuum level. For example, the local energy density
in such a state can be negative, although the total energy is always
positive. The effects of squeezed states of photons on coherent electrons
were recently analyzed by us\cite{HF08}. In this section, we will give
a summary of the results found there. 
 Consider the special case where the quantized electromagnetic field
is in a state in which one mode is excited to a squeezed vacuum state, 
and all other
modes remain in the ground state. We take the excited mode to be a
plane wave in a box with periodic boundary conditions,
 with wave vector $\bar{\mathbf{k}}$ and polarization $\bar{\lambda}$,
 so the quantum state may be denoted by 
$|\zeta_{\bar{\lambda}\bar{\mathbf{k}}} \rangle$. This is a one complex parameter
family of states, labeled by 
$\zeta_{\bar{\lambda}\bar{\mathbf{k}}}=r\,{\rm e}^{i\theta}$. 
Take the plane wave 
to be travelling in the $y$-direction, with linear polarization in the
$z$-direction. For electrons emitted at $t=t_0$, we take the paths $C_1$
and $C_2$ to be given by
\begin{equation}
z(t) =\pm \frac{R}{T^4}\,[(t-t_0 -T)^2 -T^2]^2 \,.
\end{equation}
That is, the electrons start at $z=0$ at $t=t_0$, reach their maximum
separations where $z=\pm R$ at $t=t_0 +T$, and finally return to $z=0$
at $t=t_0+2T$. Here $2T$ and $2R$ can be thought of as the effective flight 
time and path separation, respectively. Electrons which start from the source 
at different times will experience different fluctuations. 

We are interested in the change in the coherence functional due to the 
non-trivial
state of the photon field, so define a renormalized coherence functional
$W_R = W -W_0$, where $W_0$ is the functional in the Minkowski vacuum state.
In our case, one finds that
\begin{equation}
    W_R=-\frac{8\pi\alpha}{V\bar{\omega}}M\,g\left(r,t_0\right)\,,
\end{equation}
where $\alpha$ is the fine structure constant, $V$ is the normalization volume,
and $\bar{\omega}$ is the frequency of the excited photon mode. In addition,
\begin{equation}
	 M=\left(\frac{16R}{\bar{\omega}^4T^4}\right)^2
\Bigl[(-3+\bar{\omega}^2T^2)\sin\bar{\omega}T+3\bar{\omega}T\cos\bar{\omega}T
\Bigr]^2\,,
\end{equation}
which  does not depend on the electron emission time $t_0$ and is always 
positive definite. Thus the sign of $W_R$ is solely determined by the quantity
\begin{equation}
    g\left(r,t_0\right)=
\eta\left[\mu\cos\left(\alpha+\beta t_0\right)+\eta\right]\,,
   \label{eq:g}
\end{equation}
where $\mu=\cosh r$, $\eta=\sinh r$, and 
\begin{equation}
    \alpha=\bar{\omega}T-\theta\,,\qquad\beta=2\bar{\omega}\,.
\end{equation}

\begin{figure}
\begin{center}
    \scalebox{1.2}{
        \includegraphics{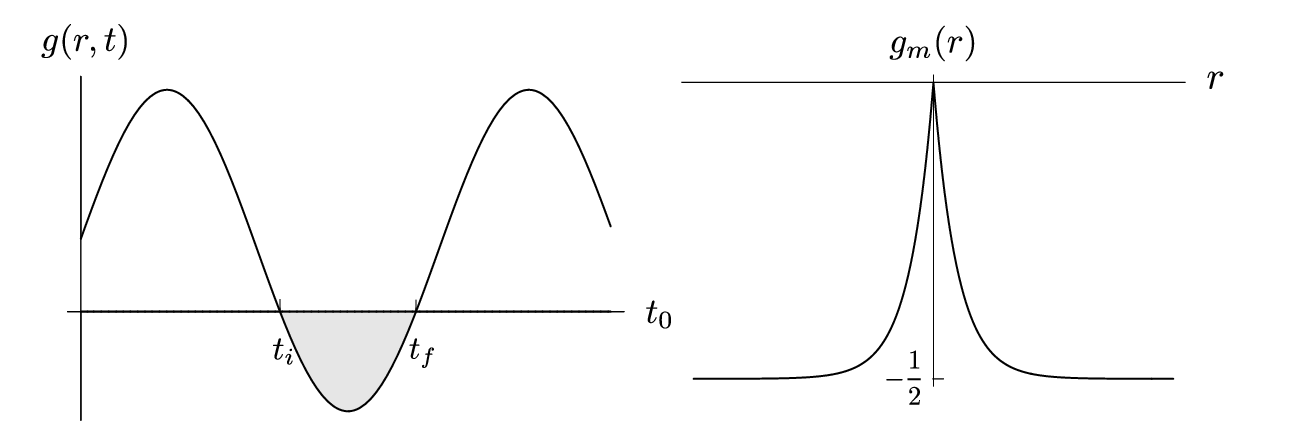}}
\caption{The left figure shows the behavior of $g\left(r,t\right)$ defined in
     Eq.~(\ref{eq:g}), as a function of the emissions time
 $t_0$. The right figure shows how how the minimum value of $g$ as a
     function of $t_0$, $g_m\left(r\right)$, depends on $r$.}
\label{Fi:squeezed_state}
\end{center}
\end{figure}

The behavior of $g(r,t_0)$ as a function of $t_0$ for fixed $r$ is illustrated 
in the left part of Fig.~\ref{Fi:squeezed_state}. The key feature is that
  $g(r,t_0) <0$ for $t_i<t_0<t_f$. This means that for electrons emitted
during this interval, we have $W_R >0$, an increase of contrast compared
to the photon vacuum state. This is the phenomenon of {\it recoherence}.
The minimum value of $g$ in the interval  $t_i<t_0<t_f$ is $g_m(r)$, plotted
in the right part of Fig.~\ref{Fi:squeezed_state}, form which we see that
 $g(r,t_0) > -\frac{1}{2}$.

\begin{figure}
\begin{center}
    \scalebox{1.2}{
        \includegraphics{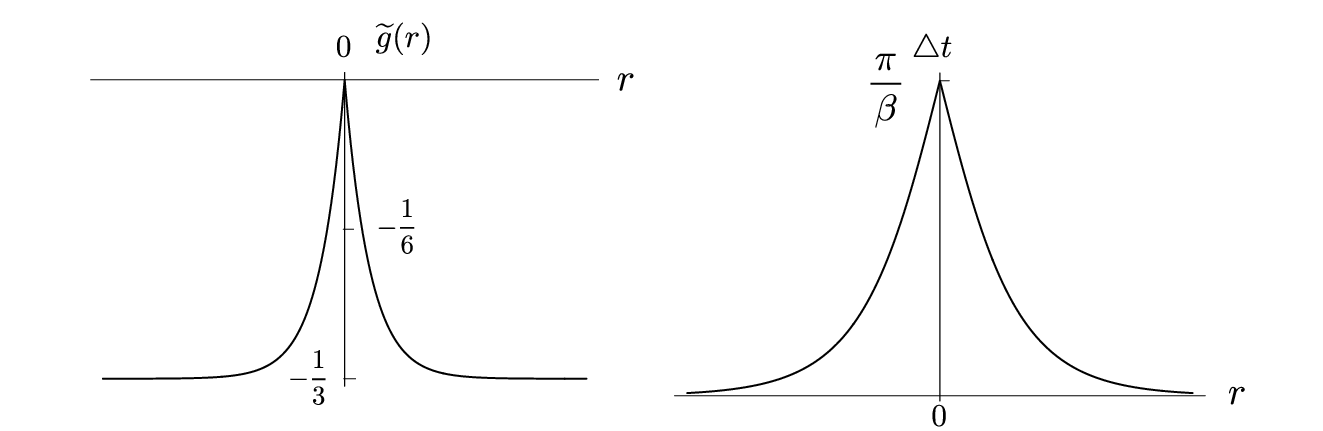}}
\caption{The left figure shows how $\widetilde{g}\left(r\right)$, the average of
$g(r,t)$ over the interval when $g(r,t) < 0$, as a function of $r$. 
The right figure illustrates the width of this interval, $\triangle
t$, also as a function of $r$.}
\label{Fi:squeezed_state1}
\end{center}
\end{figure}

Let $\widetilde{g}\left(r\right)$ and $\widetilde{W}_R$ denote the averages
of  $g(r,t_0)$ and of $W_R$, respectively, over the interval $t_i<t_0<t_f$.
The dependence of $\widetilde{g}\left(r\right)$ upon $r$ is illustrated
in Fig.~\ref{Fi:squeezed_state1}, where we see that  
$\widetilde{g}\left(r\right) > -\frac{1}{3}$. This bound limits the degree 
of recoherence, and is analogous to the quantum inequalities 
which limit the magnitude and duration of negative energy densities in
quantum field theory\cite{lF1978,Ford91,FR97,FE98}.
Marecki\cite{Marecki02,Marecki08} has recently derived
variants of the quantum inequalities for limiting the amount of
squeezing which might be observed in photodetection experiments in
quantum optics. The bounds on  $\widetilde{g}\left(r\right)$ and 
on $ \widetilde{W}_R$ are sufficient to ensure that unitarity is always
preserved, so that $W<0$. Thus the recoherence effect can never be greater
in magnitude that the decoherence in the vacuum state.  

The bound given above may be used to make estimates of the maximum recoherence.
If we assume that  $ \widetilde{W}_R$ attains its maximum value, then for a
single mode in a box, we find
\begin{equation}
\widetilde{W}_R \approx 8 \times 10^{-4}\, \frac{\lambda^3}{V} \,
 \left(\frac{R}{T}\right)^2\,
 \left(\frac{\lambda}{T}\right)^2\,,
\end{equation}
where $\lambda = 2\pi/\bar{\omega}$ is the wavelength of the excited mode.
This estimate was derived assuming non-relativistic motion and periodic
boundary conditions. However, it should serve as a rough estimate for
more realistic cavities. 
For a rough estimate, let us take $V \approx \lambda^3$ and 
$R \approx \lambda$, corresponding to
the lowest frequency mode in the cavity and a path separation 
of the order of the cavity size.
This leads to
\begin{equation}
\widetilde{W}_R \approx 10^{-3} \left(\frac{R}{T}\right)^4\,.
                   \label{eq:Wave}
\end{equation}
Non-relativistic motion requires $T \gg R$. If, for example, we take
$R/T \approx 1/10$, we would get the estimate  $\widetilde{W}_R
\approx 10^{-7}$. 

The case where a finite bandwidth of modes is excited is treated in 
our previous paper\cite{HF08}. If $\Delta \omega$ is the bandwidth 
of excited modes
and $\Delta \Omega$ is the solid angle within which they lie, then
one finds the estimate
\begin{equation}
\widetilde{W}_R \approx  10^{-2}\,  \left(\frac{R}{T}\right)^2\,
 \frac{\Delta\omega}{\bar{\omega}}\,   \Delta\Omega \,.
\end{equation}
All of the factors in the above expression, $R/T$,
 ${\Delta\omega}/{\bar{\omega}}$, and
$\Delta \Omega$, should be small compared to unity for our analysis to
be strictly valid. If we take all three of these factors to be of
order $10^{-1}$, then we would obtain  $\widetilde{W}_R \approx
10^{-6}$.

\section{Summary}
In this article, we have reviewed selected aspects of decoherence, using
coherent electrons coupled to the quantized electromagnetic field as
our model quantum system. This system exhibits all three forms of decoherence,
dephasing, entanglement, and ``which-path'' information. The basic effect
of the photon field can be described either as photon emission or as
Aharonov-Bohm phase fluctuations. If the photon field is not in its vacuum
state, then the decoherence is typically larger than in the vacuum.
However, we have discussed how this can be reversed for selected quantum
states, resulting in recoherence. The phenomenon of recoherence is a
sub-vacuum phenomenon, similar to quantum violations of the weak energy
condition. Both effects are limited in magnitude by quantum inequalities.
Although the effect of recoherence is small, its eventual measurement is
a possibility.

\section*{Acknowledgments}
We would like to thank C.I. Kuo, D.S. Lee, K.W. Ng and C.H. Wu for 
useful discussions. This  work was supported in part by the National 
Science Foundation under Grant PHY-0555754.

\end{document}